\documentclass[onecolumn,amsmath,amssymb]{revtex4}

\usepackage{graphicx}
\usepackage{dcolumn}
\usepackage{bm}

\begin{document}

\preprint{APS/123-QED}

\title{Theoretical investigation of synchronous totally asymmetric exclusion processes \\on lattices with multi-input-single-output junctions}

\author{Ruili Wang$^{\ast\S}$, Mingzhe Liu$^{\ast}$ and Rui Jiang$^{\dagger}$ 
}
\address{$^{\ast}$School of Engineering and Advanced Technology, Massey University, New Zealand}
\address{$^{\S}$ State Key Laboratory for Novel Software Technology, Nanjing
University, China}
\address{$^{\dagger}$School of Engineering Science, University of Science and Technology of China, Hefei 230026, China}%

\date{\today}

\begin{abstract}
In this paper, we investigate the dynamics of synchronous totally
asymmetric exclusion processes (TASEPs) on lattices with a
multi-input-single-output (MISO) junction, which consists of $m$
subchains for the input and one main chain for the output. An MISO
junction is a type of complex geometry, which is relevant to many
biological processes as well as vehicular and pedestrian traffic
flow. A mean-field approach is developed to deal with the junction
that connects the subchains and the main chain. Theoretical
calculations for stationary particle currents, density profiles and
a phase diagram have been obtained. It is found that the phase
boundary moves toward the left in the phase diagram with the
increase of the number of subchains. The non-equilibrium stationary
states, stationary-state phases and phase boundary are determined by
the boundary conditions of the system as well as by the number of
subchains. The density profiles obtained from computer simulations
show a very good agreement with our theoretical analysis.

PACS numbers:  05.70.Ln, 02.50.Ey, 05.60.Cd

\end{abstract}

\maketitle

\section{\label{sec:Intro}Introduction}
Non-equilibrium transport phenomena have attracted much attention of
physicists since physical principles underlying these phenomena
could be revealed in terms of phases and phase transitions
\cite{HELBING01,CHOWDHURY05}. Totally asymmetric simple exclusion
processes (TASEPs) serve as a basic model for non-equilibrium
systems and have been widely applied in the study of transport
phenomena in chemistry, physics and biology, for example, particle
diffusion through membrane channels \cite{CHOU98}, the kinetics of
synthesis of proteins \cite{SHAW03}, polymer dynamics in dense media
\cite{SCH99}, gel electrophoresis \cite{WIDOM91}, intracellular
transport of motor proteins moving along cytoskeletal filaments
\cite{KLUMPP03}, vehicular traffic \cite{CHO00,HELBING01} and ant
traffic \cite{JOHN04}.

TASEPs are non-equilibrium one-dimensional lattice models in which
particles move along one direction with hard-core interactions. The
exact solution of TASEPs has been obtained by using the matrix
product ansatz (MPA) \cite{DER98} and the Bethe ansatz \cite{SCH01},
respectively. Recently, some extensions of TASEPs focus on coupling
with Langmuir kinetics
\cite{LIPOWSKY01,LIPOWSKY03,PAR03,POPKOV03,MIRIN03,EVANS03,JUH04,MUK05,NIS05,MIT06}
and/or lattice geometries
\cite{PRO04,PRO05,MIT05,STUKALIN06,REI06,JIANG07,WANG07}, as well as
bottleneck-induced transport phenomena such as in
\cite{KLU04,PIE06}. Ref. \cite{KLU04} investigated diffusive
compartments as bottlenecks in a driven transport. They found that
when a diffusive bottleneck is at the boundary of a system, the
system cannot reach a maximal current phase; when a diffusive
bottleneck is in the interior of the system, the system is dictated
by the diffusive bottleneck and has a maximal current defined by the
bottleneck. More recently, Ref. \cite {PIE06} introduced a
bottleneck phase to describe the phenomenon that the current is
independent of boundary conditions.

In this paper, we focus on one special lattice geometry - junctions,
which are widely observed in many real physical systems. Those
junctions are formed for various reasons. For example, (i) variation
of the number of protofilaments on a microtubule \emph{in vitro}
\cite{CHRETIEN92}; (ii) transport of vesicles in a branching axon or
dendrite \cite{BURACK00}; (iii) merging of two or more roads; (iv)
data through hubs (e.g., switches, routers) on the Internet. This
kind of lattice geometry can cause congestion, e.g., in the traffic
of molecular motors, vehicles or data packets. Traffic congestion of
molecular motors could lead to some human diseases such as
Alzheimers \cite{GOLDSTEIN01} and some neurodegenerative diseases
\cite{HURD96}. Vehicular traffic congestion can pollute environment,
increase fuel consumption.

Inspired by this wide range of possible applications, we investigate
the dynamics of synchronous (i.e., in \emph{parallel update})
totally asymmetric exclusion processes on lattices with a
multi-input-single-output (MISO) junction (see Fig. 1). The parallel
updating procedure has been typically adopted in modeling vehicular
and pedestrian traffic \cite{CHO00,HELBING01}. Multiple inputs
exhibit more complex interactions between particles at junction
points than two inputs. In reality, it can be observed that several
traffic lanes merge into one lane and multiple protofilaments come
together to form one protofilament \cite{CHRETIEN92}. However, they
have not been understood well from the viewpoint of theoretical
analysis.

Theoretical calculations, along with a mean-field approximation, are
developed. The phase diagram is presented and density profiles are
investigated. A phenomenological domain wall theory, based on Refs.
\cite{ABK98,PRO05}, is used to predict phase boundaries. Computer
simulations are also conducted. Note that TASEPs on lattices with
Y-junctions (e.g., two-input-single-out junctions) in a \emph{random
updating} procedure has been investigated in Ref. \cite{PRO05}.
Y-junctions can be seen as a special case of
multi-input-single-output junctions.

The paper is organized as follows. In Section II, we give a
description of a synchronous TASEP model with an MISO junction,
theoretical calculations as well as the mean-field approximation are
developed. In Section III, we analyze the phase boundaries using a
phenomenological domain wall theory. In Section IV, the results of
our theoretical calculations and computer simulations are presented.
Finally, we give our conclusions in Section V.

\section{\label{sec:Model}The Model and Mean-field Approximation}

An MISO junction is illustrated in Fig. 1. The system consists of
$m$ subchains for input and one main chain (chain $m+1$) for output
connected by junction points--sites $N$ on the subchains and site
$N+1$ on the main chain. Each subchain and the main chain includes
$N$ sites. For simplification, inter-lane transitions between
subchains are not permitted in this paper. Particles are assumed to
move from the left to the right.

\begin{figure}[!h]
\includegraphics[width= 3 in, height=2 in]{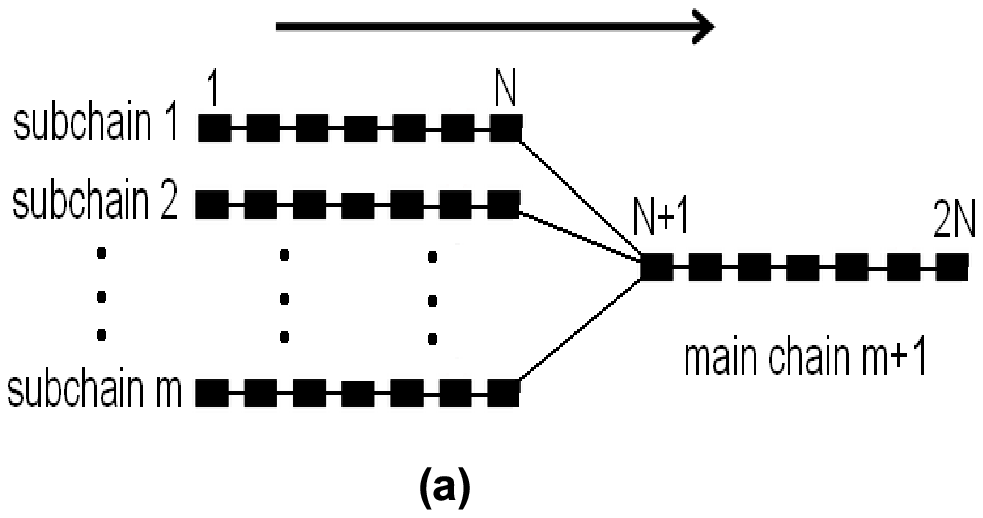}
\includegraphics[width= 3 in, height=2 in]{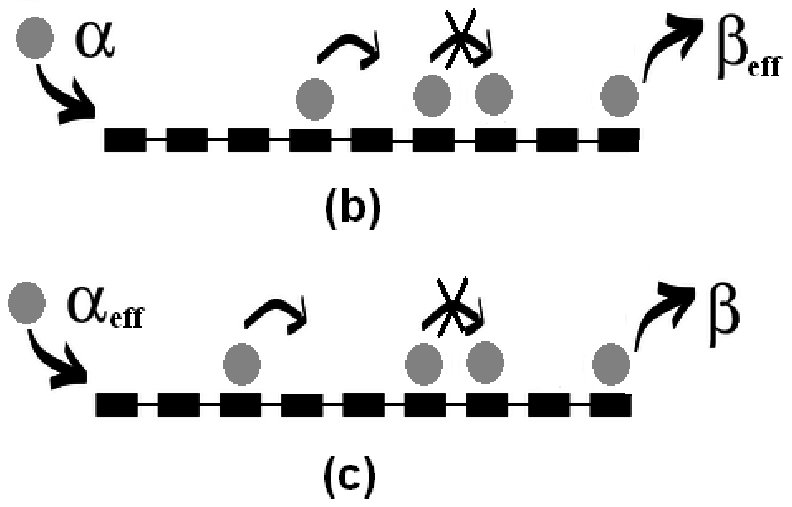}
\caption{\label{fig:FundDiag} (a) Schematic diagram of a synchronous
TASEP with an MISO junction. Particles move from the left to the
right with hard-core exclusion. (b) In a subchain, the injection
rate at site 1 and the ejection rate at site $N$ are given by
$\alpha$ and $\beta_{eff}$, respectively. (c) In the main chain, the
injection rate at site $N+1$ and the ejection rate at site $2N$ are
given by $\alpha_{eff}$ and $\beta$, respectively.}
\end{figure}

An occupation variable, $\tau_{\ell,i}$, denotes the state of the
$i$th site in the $\ell$th subchain ($\ell=1,2,...,m$) and the main
chain ($\ell= m+1$). $\tau_{\ell,i}=1$ (or $\tau_{\ell,i}=0$) means
that site $\tau_{\ell,i}$ is occupied (or empty). The system updates
all particles \emph{in parallel} by the following rules (see Fig.
1):

\begin{itemize}
\item $i=1$. (i) If $\tau_{\ell,1}=0$, a particle enters the system at rate
$\alpha$; or (ii) If $\tau_{\ell,1}=1$ and $\tau_{\ell,2}=0$, then
the particle at site $(\ell,1)$ moves into site $(\ell,2)$; or (iii)
If both $\tau_{\ell,1}=1$ and $\tau_{\ell,2}=1$, then the particle
at site $(\ell,1)$  does not move.
\item $i=N$. (i) If sites $N$ of $k$ subchains $(1<k\leq m)$ are occupied by $k$ particles at the same time,
particles have the same priority to hop to site $N+1$. However, only
one particle will enter site $N+1$ at any single time step,
providing that site $N+1$ is empty; or (ii) If only one site $N$ of
the subchains is occupied by a particle, the particle can directly
hop to site $N+1$ providing that site $N+1$ is empty.
\item $i=2N$. If $\tau_{m+1,2N}=1$, the particle leaves the system with rate $\beta$.
\item $1<i<N$ or $N+1\leq i<2N$. If $\tau_{\ell,i}=1$, the particle can move into site $(\ell,i+1)$ providing
$\tau_{\ell,i+1}=0$. Otherwise, the particle cannot move.
\end{itemize}

Exactly solvable results of an one-dimensional synchronous TASEP
have been obtained in Ref. \cite{TIL98,GIER99}. We briefly recall
these results, as the solution of our synchronous TASEP with a
junction can be derived from them. There are three phases (low
density (LD), high density (HD) and maximal current(MC)) and a
transition line when $\alpha=\beta$.
The MC $J = 0.5$ can only be reached at $\alpha=\beta=1$
\cite{GIER99}.

\begin{itemize}
\item When $\alpha< \beta\leq1$, a low-density (LD) phase is obtained with
\begin{equation}
J=\rho, \quad\rho=\rho_1, \quad \rho_1=\frac{\alpha}{1+\alpha},
\quad\rho_N=\frac{\alpha}{\beta(1+a)}.
\end{equation}
where $J$ is the system current; $\rho$ is the bulk density;
$\rho_1$ ($\rho_N$) is the particle density at the first (last)
site.

\item When $1\geq\alpha> \beta$, a high-density (HD) phase is obtained
with
\begin{equation}
J=1-\rho, \quad\rho=\frac{1}{1+\beta}, \quad
\rho_1=1-\frac{\beta}{\alpha(1+\beta)},\quad \rho_N=\rho.
\end{equation}

\item When $\alpha = \beta <1$, a transition line
between LD and HD is obtained.

\item When $\alpha = \beta =1$, the
maximal current (MC) is obtained and $J = 0.5$.

\end{itemize}

Based on the above results, we develop exactly solvable results for
TASEPs with an MISO junction.  For an MISO junction, as the current
is conserved through the system, we have:
\begin{equation}
J_1+J_2+\cdot\cdot\cdot+J_m=J_{m+1}, \quad
J_1=J_2=\cdot\cdot\cdot=J_m, \quad m J_ \ell=J_{m+1}\leq 0.5
\end{equation}
where $J_{\ell}$ ($\ell=1,2,...,m$) is the current on the $\ell$th
subchain; $J_{m+1}$ is the current of the main chain.

Each of the $m$ subchains of the MISO junction can be seen as a
synchronous TASEP with injection rate $\alpha$ and ejection rate
$\beta_{eff}$, while the main chain can be seen as a synchronous
TASEP with injection rate $\alpha_{eff}$ and ejection rate $\beta$.
$\alpha_{eff}$ and   $\beta_{eff}$  can be written as
\begin{equation}
\beta_{eff}=1-\rho_{N+1}, \quad \alpha_{eff}=m\rho_N.
\end{equation}

These $m$ subchains should have the identical phases when particles
on the $m$ subchains merge into the main chain with the same
priority. Our computer simulations also support this prediction.
Thus, the stationary state of the system can be obtained by
combining the possible phases that exist in each of these subchains
and the main chain. As each single chain may have three possible
phases (LD, HD and MC), due to the equivalence of these subchains,
the number of possible stationary phases of the system is equal to
$3^2=9$. In other words, a stationary state can be one of the
following nine phases: the (LD, LD), (LD, HD), (LD, MC), (HD, LD),
(HD, HD), (HD, MC), (MC, LD), (MC, HD), and (MC, MC) phases.

One can see that three phases cannot exist: (MC, LD), (MC, HD) and
(MC, MC). According to Eq. (3), it is impossible for the maximal
current phase to exist in a subchain since the maximal possible
current in the system is no more than 0.5. Therefore, the number of
the possible phase combinations reduce to 6, i.e., the (LD, LD),
(LD, HD), (LD, MC), (HD, LD), (HD, HD), (HD, MC) phases.

\begin{itemize}
\item The (LD, HD) phase. The conditions
for this case are as follows:

\begin{equation}
\alpha< \beta_{eff}, \quad\alpha_{eff}> \beta.
\end{equation}

From Eqs. (1) and (2), the stationary properties of this phase are
given by:

\begin{displaymath}
J_1=\frac{\alpha}{1+\alpha}, \quad J_{m+1}=\frac{\beta}{1+\beta},
\quad \rho_1=\frac{\alpha}{1+\alpha},
\end{displaymath}
\begin{equation}
\rho_N=\frac{\alpha}{\beta_{eff}(1+\alpha)}, \quad
\rho_{N+1}=1-\frac{\beta}{\alpha_{eff}(1+\beta)}, \quad
\rho_{2N}=\frac{1}{1+\beta}.
\end{equation}

According to Eq. (3), $mJ_1=J_{m+1}$, we get:
\begin{equation}
\alpha=\frac{\beta}{m+(m-1)\beta}
\end{equation}

However, $\alpha_{eff}$ and $\beta_{eff}$ are not solvable from
above equations. In other words, we cannot calculate the bulk
density through the above equations. The density will be solved
through the domain wall theory in Section III. From Figs. 2(a) and
(b), one can see that $\alpha=\beta/[m+(m-1)\beta]$ (when $\beta <
1$) corresponds to the transition line between the (LD, LD) phase
and the (HD, HD) phase.

\item The (LD, MC) phase. This phase corresponds to the following conditions:

\begin{equation}
\alpha< \beta_{eff}, \quad\alpha_{eff}=\beta=1.
\end{equation}
According to Eqs. (1) and (3), we obtain:
\begin{equation}
\alpha=\frac{1}{2m-1}, \quad J_1=\frac{1}{2m},\quad J_{m+1}= 0.5.
\end{equation}

Obviously, $\alpha=1/(2m-1)<1/2$ when $m\geq 2$. That is, Eq. (8) is
satisfied. Thus, the (LD, MC) phase can exist in the system when:
\begin{equation}
\alpha=\frac{1}{2m-1}, \quad\beta=1.
\end{equation}

Again, we cannot calculate the bulk density through above equations.
The density will also be solved through domain wall theory in
Section III. From Figs. 2(a) and (b), one can see that the (LD, MC)
phase is the transition phase between the (LD, LD), (LD, HD), (HD,
HD) and (HD, MC) phases.

\item  The (LD, LD) phase. The following conditions should be
satisfied:
\begin{equation}
\alpha<\beta_{eff}, \quad \alpha_{eff}<\beta.
\end{equation}

According to Eq. (1), the stationary current and density are given
by:

\begin{displaymath}
J_1=\frac{\alpha}{1+\alpha}, \quad
J_{m+1}=\frac{\alpha_{eff}}{1+\alpha_{eff}}, \quad
\rho_1=\frac{\alpha}{1+\alpha},
\end{displaymath}
\begin{equation}
\rho_N=\frac{\alpha}{\beta_{eff}(1+\alpha)}, \quad
\rho_{N+1}=\frac{\alpha_{eff}}{1+\alpha_{eff}}, \quad
\rho_{2N}=\frac{\alpha_{eff}}{\beta(1+\alpha_{eff})}.
\end{equation}

Using Eqs. (3) and (4), $\alpha_{eff}$ and $\beta_{eff}$ are
expressed as:

\begin{equation}
\alpha_{eff}=\frac{m\alpha}{1-(m-1)\alpha}, \quad
\beta_{eff}=\frac{1-(m-1)\alpha}{1+\alpha}.
\end{equation}

Since $ \alpha_{eff}\leq 1$ and
$\alpha_{eff}=m\alpha/[1-(m-1)\alpha]$, $\alpha\leq 1/(2m-1)$.
Substituting Eq. (13) into Eq. (11), we obtain $\alpha<\sqrt{1+
m^2/4}-m/2$ for $\alpha<\beta_{eff}$, and
$\alpha<\beta/[m+(m-1)\beta]$ for $\alpha_{eff}<\beta$. Since
$1/(2m-1)<\sqrt{1+m^2/4}-m/2$ (when $m\geq 2$) and
$\beta/[m+(m-1)\beta]\leq 1/(2m-1)$ (as $\beta\leq 1$), the system
is in the (LD, LD) phase when:

\begin{equation}
\alpha<\frac{\beta}{m+(m-1)\beta},\quad \beta\leq 1.
\end{equation}

\item The (HD, HD) phase. The conditions for this case are given by:

\begin{equation}
\alpha> \beta_{eff}, \quad \alpha_{eff}>\beta.
\end{equation}

The current and density of this phase in a stationary state are:
\begin{displaymath}
J_1=\frac{\beta_{eff}}{1+\beta_{eff}}, \quad
J_{m+1}=\frac{\beta}{1+\beta}, \quad
\rho_1=1-\frac{\beta_{eff}}{\alpha(1+\beta_{eff})},
\end{displaymath}
\begin{equation} \label{eq:HD,HD}
\rho_N=\frac{1}{1+\beta_{eff}}, \quad
\rho_{N+1}=1-\frac{\beta}{\alpha_{eff}(1+\beta)},
\quad\rho_{2N}=\frac{1}{1+\beta}.
\end{equation}

From Eqs. (3) and (\ref{eq:HD,HD}), we obtain
$\beta_{eff}=\beta/[m+(m-1)\beta]$. Thus, the system is in the (HD,
HD) phase when:
\begin{equation}\label{eq:HD.HD}
\alpha>\frac{\beta}{m+(m-1)\beta} 
\end{equation}

\item The (HD, MC) phase. The corresponding conditions for this phase are:

\begin{equation}
\alpha> \beta_{eff}, \quad\alpha_{eff}=\beta=1.
\end{equation}
According to Eqs. (2-3), we obtain
\begin{equation}
J_1=\frac{1}{2m},\quad\rho_N=\frac{1}{m},
\quad\rho_{N+1}=\frac{2m-2}{2m-1}, \quad\beta_{eff}=\frac{1}{2m-1}.
\end{equation}

Thus, the (HD, MC) phase can exist in the system when:
\begin{equation}\label{eq:HD.MC}
\alpha>\frac{1}{2m-1}, \quad\beta=1.
\end{equation}

\item (HD, LD) phase. The conditions of existence for this phase can be written as

\begin{equation}
\alpha> \beta_{eff}, \alpha_{eff}<\beta.
\end{equation}

The corresponding expressions for stationary current and density are

\begin{displaymath}
J_1=\frac{\beta_{eff}}{1+\beta_{eff}}, \quad
J_{m+1}=\frac{\alpha_{eff}}{1+\alpha_{eff}}, \quad
\rho_1=1-\frac{\beta_{eff}}{\alpha(1+\beta_{eff})},
\end{displaymath}
\begin{equation}
\rho_N=\frac{1}{1+\beta_{eff}}, \quad
\rho_{N+1}=\frac{\alpha_{eff}}{1+\alpha_{eff}}, \quad
\rho_{2N}=\frac{\alpha_{eff}}{\beta(1+\alpha_{eff})}.
\end{equation}

According to Eq. (3), we have
\begin{equation}
\alpha_{eff}=\frac{m\beta_{eff}}{1-(m-1)\beta_{eff}}.
\end{equation}

From Eqs. (4) and (22), $\alpha_{eff}$ and $\beta_{eff}$ can be
rewritten as follows
\begin{equation}
\alpha_{eff}=\frac{m}{1+\beta_{eff}}, \quad
\beta_{eff}=\frac{1}{1+\alpha_{eff}}.
\end{equation}
\end{itemize}

Substituting Eq. (24) into Eq. (23), we obtain
$\alpha_{eff}=\sqrt{1+m^2/4}+m/2-1$ and
$\beta_{eff}=\sqrt{1+m^2/4}-m/2$. It can be seen that values of
$\alpha_{eff}$ and $\beta_{eff}$ are determined by the number of
subchains $m$, independent of $\alpha$ and $\beta$. This indicates
that the (HD, LD) phase cannot be represented in the $\alpha-\beta$
plane. In other words, the (HD, LD) phase does not exist in the
system. In fact, when the subchains are in the high density phase,
particles at site $N$ will hop to site $N+1$ at almost each time
step, which leads to $\alpha_{eff}\approx 1$. Thus, it is impossible
for the main chain to maintain the low density phase.

\begin{figure}[!h]
\begin{center}
\includegraphics[width= 2.5 in, height=2.5 in]{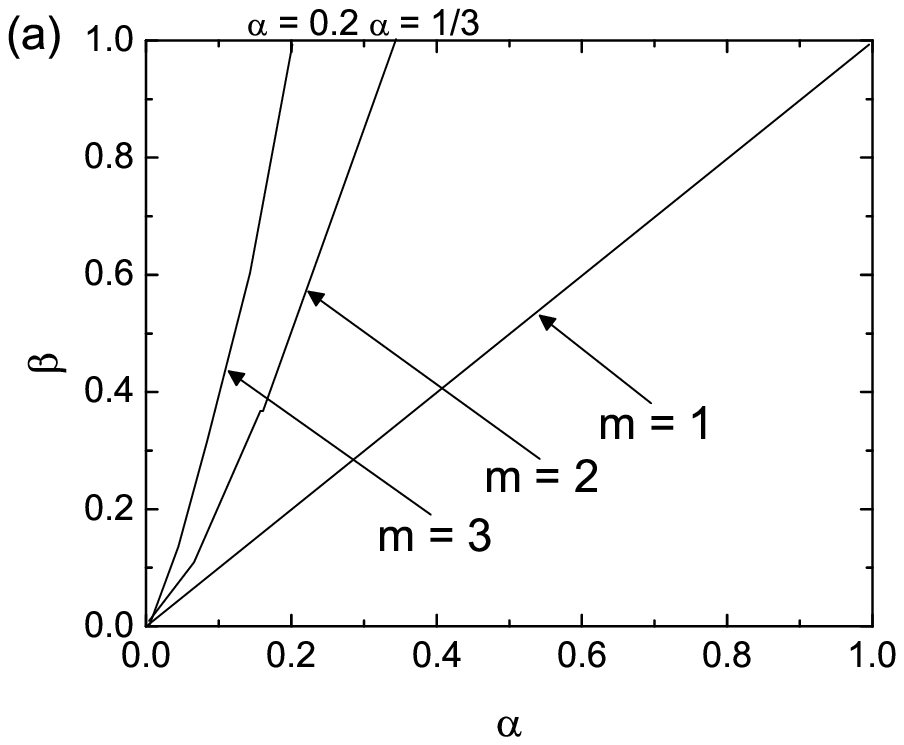}\quad
\includegraphics[width= 2.5 in, height=2.5 in]{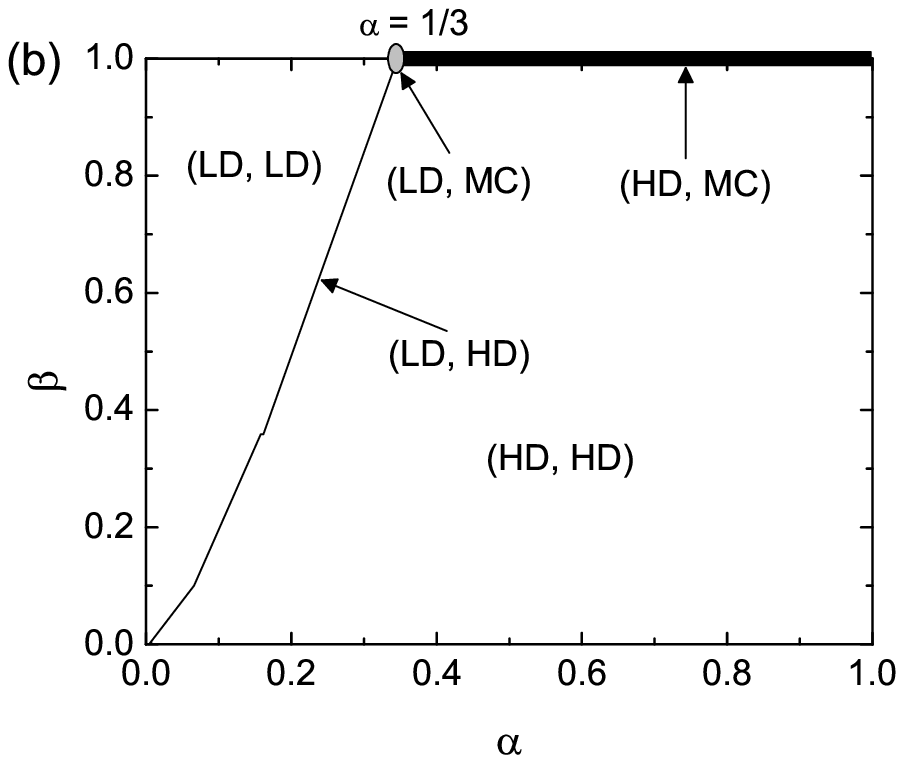}
\caption{(a) Phase boundaries (or transition lines) for $m=1,2$ and
$3$ in the synchronous TASEPs with an MISO junctions. (b) Phase
diagram for $m=2$ in the synchronous TASEPs with an MISO junction.
The solid line represents the (LD, HD) phase specified by
$\alpha=\beta/[m+(m-1)\beta]$ and $\beta<1$; the grey oval
corresponds to the (LD, MC) phase specified by $\alpha=1/(2m-1)$ and
$\beta=1$; and the black rectangle for the (HD, MC) phase specified
by $\alpha>1/(2m-1)$ and $\beta=1$.}
\end{center}
\end{figure}

From the analysis above, one can see that there are five possible
phases ((LD, LD), (LD, HD), (LD, MC), (HD, HD) and (HD, MC)) in this
system. Fig. 2(a) shows the possible phase boundaries
($\alpha=\beta/[m+(m-1)\beta]$) for $m=1, 2$ and $3$. With the
increase of $m$, we can predict that the phase boundary moves toward
the left in the phase diagram, which means that the low-density area
decreases while the high-density area increases. The phase diagram
for $m=2$ is also shown in Fig. 2(b). From Fig. 2(b), one can see
that: (i) The (LD, HD) phase corresponds to the transition line
between the (LD, LD) phase and the (HD, HD) phase specified by
$\alpha=\beta/[m+(m-1)\beta]$ and $0\leq\beta< 1$. (ii) The (LD, MC)
phase is the transition phase between the (LD, LD),(LD, HD), (HD,
HD) and (HD, MC) phases specified by $\alpha= 1/(2m-1)$ and $\beta =
1$ . Also, note that the transition from the (LD, LD) phase to the
(LD, HD) phase, the density change on the subchains is continuous,
while the density change on the main chain is discontinuous.
Similarly, the transition from the (LD, HD) to the (HD, HD) phases,
the density change on the subchains is discontinuous, while the
density change of the main chain is continuous. Also, for the
transition from the (LD, MC) phase to the (HD, MC) phase, the
density change on the subchains is discontinuous, while the density
profile on the main chain is unchanged.

From the above analysis, we can see the non-equilibrium stationary
states, stationary-state phases and the phase boundary are
determined by the boundary conditions of the system as well as by
the number of subchains. In other words, the dynamics of the system
is determined by its environment and its own structure.

\begin{figure}[!h]
\begin{center}
\includegraphics[width= 2.5 in, height=2.5 in]{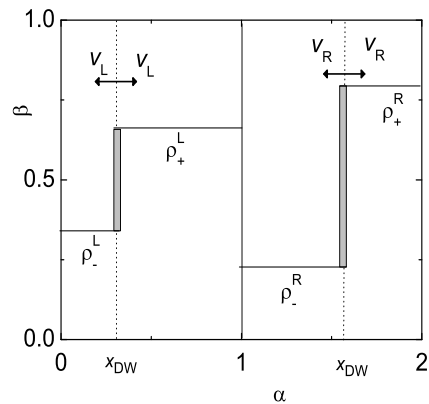}
\caption{Schematic diagram of the domain wall dynamics in the (LD,
HD) phase. The domain wall moves in the left and right subsystems at
rates $v_L$ and $v_R$, respectively.}
\end{center}
\end{figure}

\section{Domain Wall Dynamics}

A phenomenological domain wall (i.e., shock) theory to explain phase
behavior of a TASEP in a \emph{random update} procedure with open
boundaries is introduced in Ref. \cite{ABK98}. A domain wall is a
phase boundary connected by two possible stationary states. The wall
can have a random walk through the system with a drift speed defined
as follows \cite{ABK98}:
\begin{equation}\label{eq:V}
V=\frac{J_+-J_-}{\rho_+-\rho_-}.
\end{equation}
where $J$ and $\rho$ are the currents and densities in the two
phases; '+'('-') denotes the phase to the right (left) of the domain
wall. When $V> 0$, the domain wall moves to the right, while the
domain wall travels to the left when $V< 0$. For instance, if
$\alpha<\beta$ in a TASEP, the domain wall first appears at the left
end and will drift to the right later. The wall will pass through
the system, which eventually leads to the system being in a
low-density stationary state. When $V=0$, it implies that the domain
wall has no net drift between two possible phases. In this case,
stationary density profiles are linearly increased and the domain
wall can exist anywhere with equal probability in the system.

In this paper, the line specified by $\alpha=\beta/[m+(m-1)\beta]$
corresponds to the coexistence of the (LD, LD) and (HD, HD) phases.
However, the simple approximation theory may not fully reflect the
correlations near the junctions as indicated in Ref. \cite{PRO05}.
The domain wall theory is adopted in order to derive the phase
boundary and the density profile in the bulk. This theory has also
been used in Ref. \cite{PRO05}.

To determine the position of the domain wall in the system, we
define $x$ as $x = i/(2N)$, where $i$ is the site index and $2N$ is
the length of the system. In the range of $0<x\leq 1$, the domain
wall moves at rate $v_L$ in the left subsystem (the sub-chains). In
the range of $1<x\leq 2$, the domain wall moves at rate $v_R$ in the
right subsystem (the main chain), see Fig. (3).

$v_L$ and $v_R$ can be given by utilizing Eq. (\ref{eq:V}):
\begin{equation}
v_L=\frac{J_L}{\rho_+^L-\rho_-^L}, \quad
v_R=\frac{J_R}{\rho_+^R-\rho_-^R}.
\end{equation}
where:
\begin{equation}
J_L=\frac{\alpha}{1+\alpha}, \rho_+^L=\frac{1}{1+\alpha},
\rho_-^L=\frac{\alpha}{1+\alpha}, J_R=\frac{\beta}{1+\beta},
\rho_+^R=\frac{1}{1+\beta}, \rho_-^R=\frac{\beta}{1+\beta}.
\end{equation}
As a result, $v_L$ and $v_R$ are rewritten as:
\begin{equation}\label{eq:V.V}
v_L=\frac{\alpha}{1-\alpha}, \quad v_R=\frac{\beta}{1-\beta}.
\end{equation}

Similarly to Ref. \cite{PRO05}, $q_L$ ($q_R$) is denoted as a
probability to find the domain wall at any position in the left
(right) subsystem. For a special site, $i$, in the left (right)
subsystem, the probability is obviously equal to $q_L/N$ ($q_R/N$).
Then, at the junction point, we have:
\begin{equation}\label{eq:jun}
\frac{v_Lq_L}{N}=\frac{v_Rq_R}{N}.
\end{equation}

In addition, normalized $q_L$ and $q_R$ are satisfied with:
\begin{equation}\label{eq:nor}
q_L + q_R=1.
\end{equation}

Combining Eqs. (\ref{eq:jun}) and (\ref{eq:nor}), we obtain:
\begin{equation}\label{eq:q.q}
q_L=\frac{v_R}{v_L+v_R}, \quad q_R=\frac{v_L}{v_L+v_R}.
\end{equation}

According to Eq. (\ref{eq:V.V}), Eq. (\ref{eq:q.q}) implies that:
\begin{equation}\label{eq:q.q1}
q_L=\frac{\beta(1-\alpha)}{\alpha+\beta-2\alpha\beta}, \quad
q_R=\frac{\alpha(1-\beta)}{\alpha+\beta-2\alpha\beta}.
\end{equation}

Accordingly, the probabilities of the domain walls falling in
certain zones in the left and right subsystems are also given by:
\begin{equation}
Prob(x_{DW}>x)=q_Lx, \quad 0<x\leq 1,
\end{equation}
and
\begin{equation}
Prob(x_{DW}<x)=q_L+q_R(x-1), \quad 1<x\leq 2.
\end{equation}
Thus, the density at any position in the system becomes:
\begin{equation}\label{eq:p}
\rho(x)_k=\rho_-^kProb(x_{DW}>x)+\rho_+^kProb(x_{DW}<x), \quad k=L,
R
\end{equation}

Finally, from Eqs. (\ref{eq:q.q1})-(\ref{eq:p}), one can obtain:
\begin{equation}
\rho(x)_L=\frac{\alpha}{1+\alpha}+\frac{\beta(1-\alpha)^2}{(1+\alpha)(\alpha+\beta-2\alpha\beta)}x,
\quad 0<x\leq 1
\end{equation}
and
\begin{equation}
\rho(x)_R=\frac{\beta}{1+\beta}+\frac{\beta(1-\alpha)(1-\beta)}{(1+\beta)(\alpha+\beta-2\alpha\beta)}+\frac{\alpha(1-\beta)^2}{(1+\beta)(\alpha+\beta-2\alpha\beta)}(x-1),
\quad 1<x\leq 2.
\end{equation}

Densities in the boundary conditions can be calculated as
$\rho(x=0)_L=\alpha/(1+\alpha)$ and $\rho(x=2)_R=1/(1+\beta)$. These
results are completely identical with theoretical analysis in Refs.
\cite{TIL98,GIER99}. At the junction point $N$, the densities are
equal to
$\rho(x=1)_L=[\alpha^2(1-\beta)+\beta(1-\alpha)]/[(1+\alpha)(\alpha+\beta-2\alpha\beta)],
\rho(x=1)_R=\beta(1-\alpha\beta)/[(1+\beta)(\alpha+\beta-2\alpha\beta)]$.
Note that in the transition line between the (LD, LD) and (HD, HD)
phases, we obtain the relationship
 $\alpha=\beta/[m+(m-1)\beta]$.

\section{\label{sec:Results}Simulation results and discussions}

\begin{figure}[!h]
\begin{center}
\includegraphics[width= 2.5 in, height=2 in]{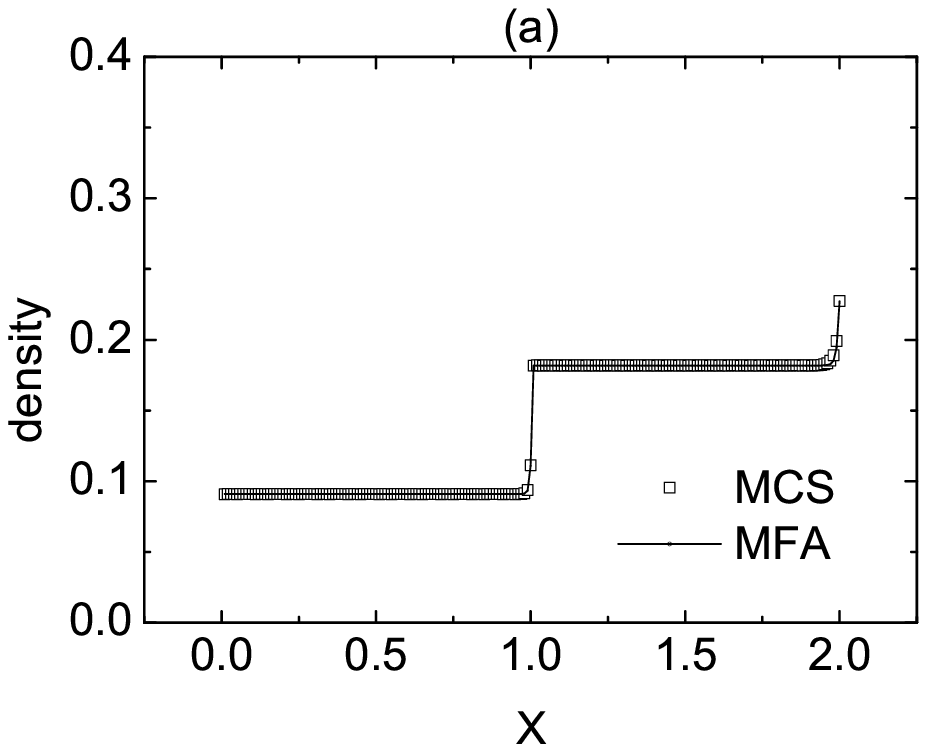} \quad
\includegraphics[width= 2.5 in, height=2 in]{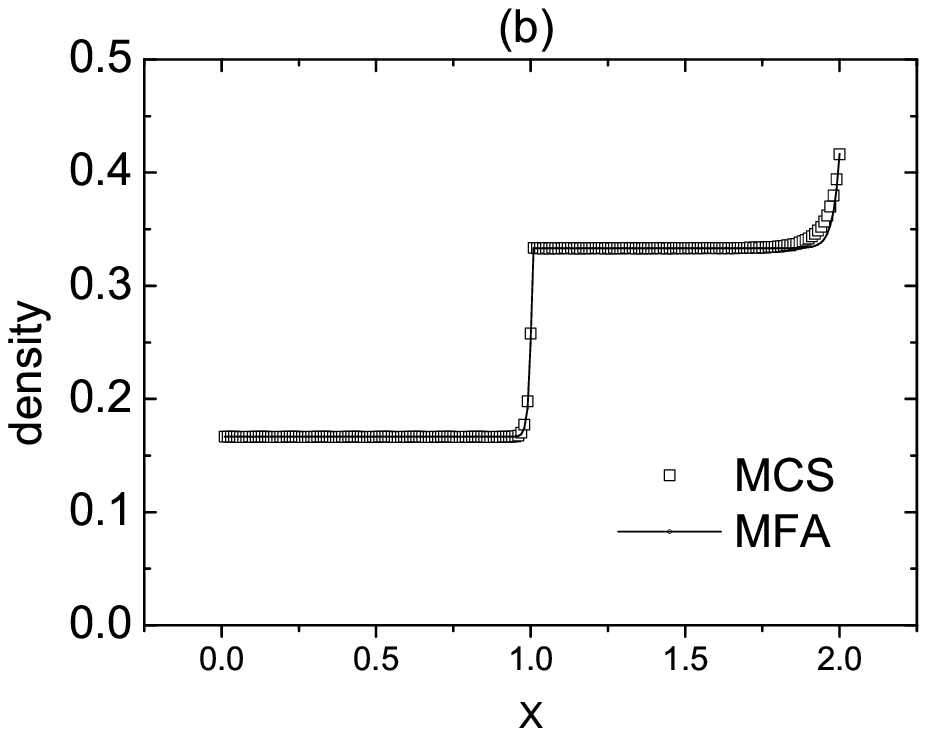} \quad
\includegraphics[width= 2.5 in, height=2 in]{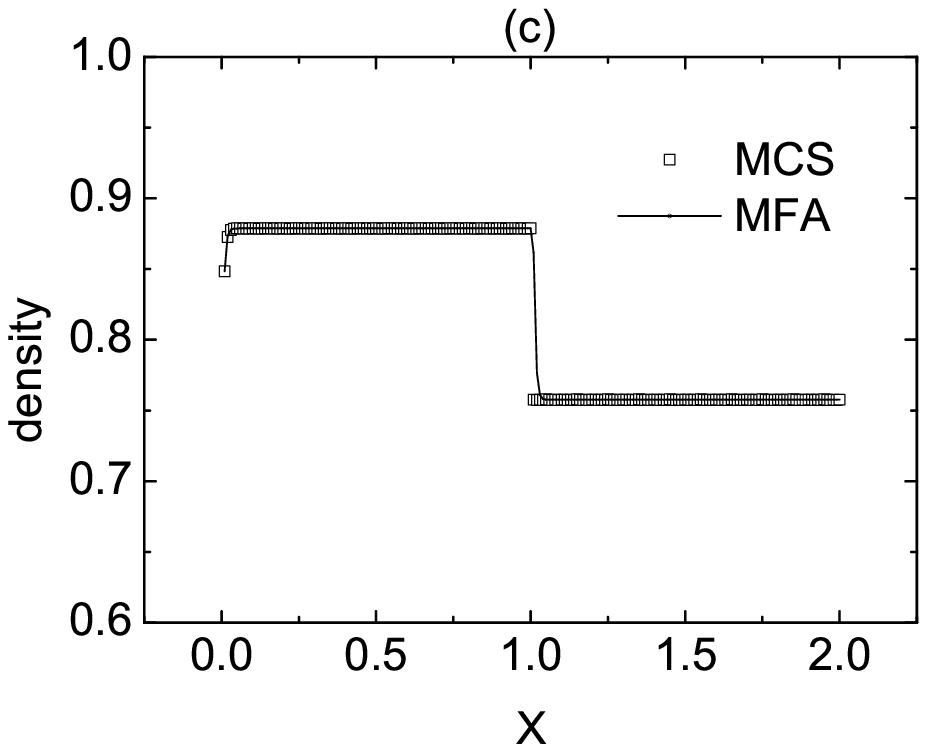} \quad
\includegraphics[width= 2.5 in, height=2 in]{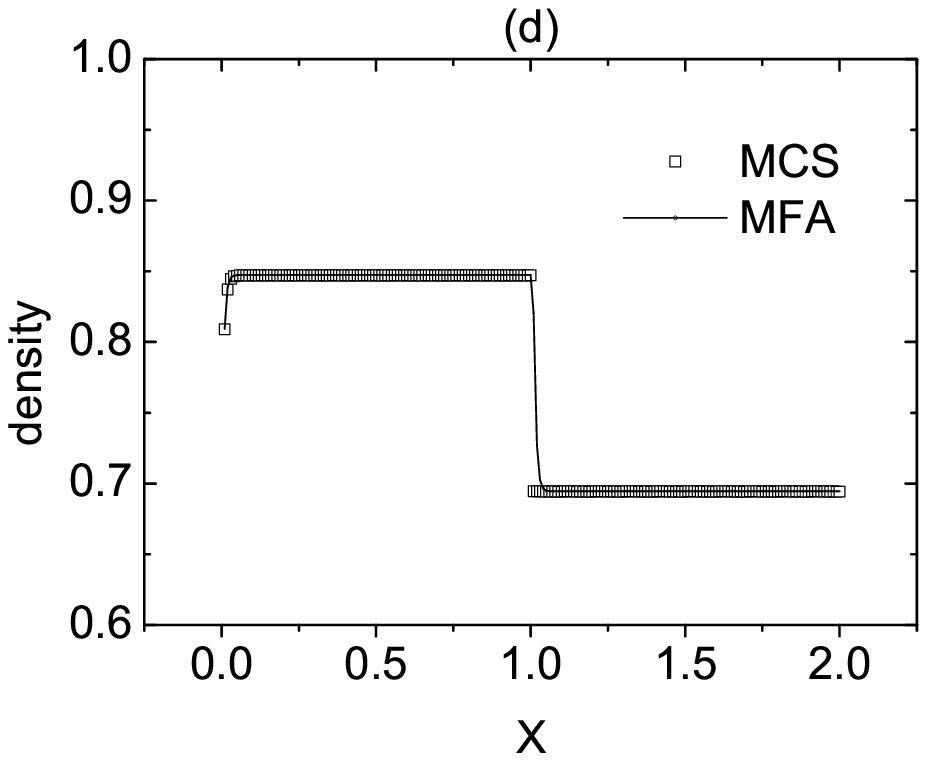} \quad
\includegraphics[width= 2.5 in, height=2 in]{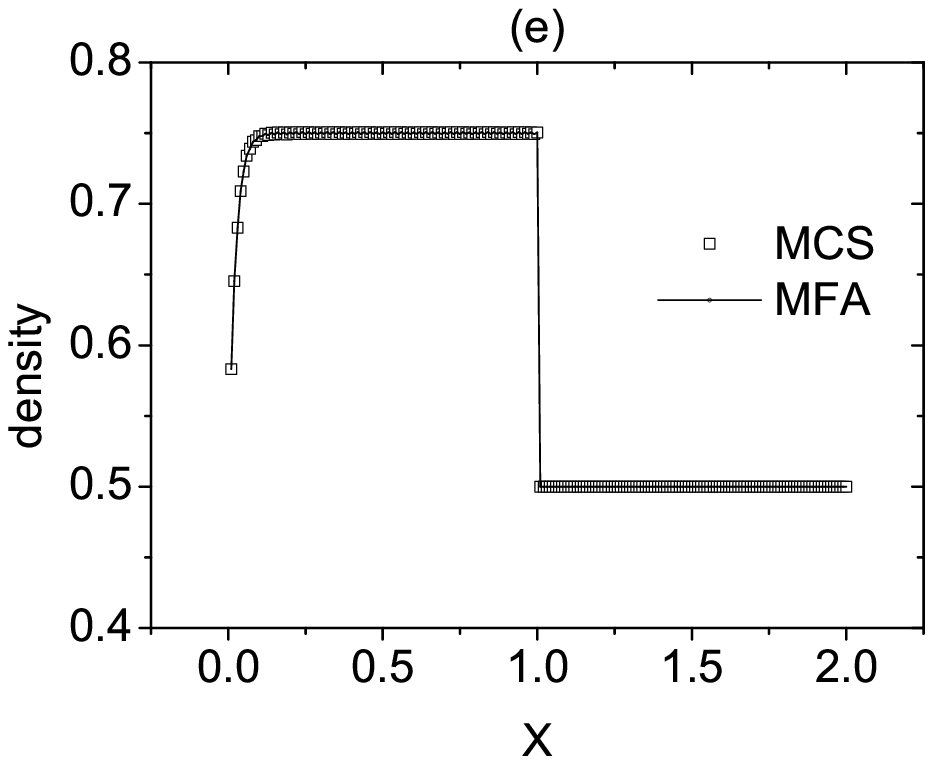}
\caption{Density profiles obtained from our theoretical calculations
and Monte Carlo simulations (MCS) when $m=2$: (a) and (b) are for
the (LD, LD) phase, (c) and (d) are for the (HD, HD) phase, and (e)
for the (HD, MC) phase. The parameters are set to: (a) $\alpha=0.1$
and $ \beta=0.8$, (b) $\alpha=0.2$ and $\beta=0.8$, (c) $\alpha=0.8$
and $ \beta=0.32$, (d) $\alpha=0.8$ and $ \beta=0.44$, and (e)
$\alpha=0.6$ and $\beta=1.0$.}
\end{center}
\end{figure}

To validate our theoretical analysis, computer simulations are
conducted. Here, we only present a synchronous TASEP with a Y-type
junction, that is $m=2$. The numbers of sites of the subchains and
the main chain are all equal to 1,000. In simulations, stationary
density profiles are obtained by averaging $10^8$ sampling at each
site. The first $10^5N$ time steps are discarded to let the
transient time out.

The density profiles for the (LD, LD), (HD, HD) and (HD, MC) phases
are shown in Fig. 4. We only illustrate the density properties of
subchain 1 and the main chain since the density properties of the
other subchain is essentially the same as subchain 1. It is found
that there is a good agreement between Monte Carlo simulations (MCS)
and mean field (MF) analysis (see Figs. 4(a)-(e)), which verifies
our theoretical investigations.

\begin{figure}[!h]
\begin{center}
\includegraphics[width= 2.5 in, height=2 in]{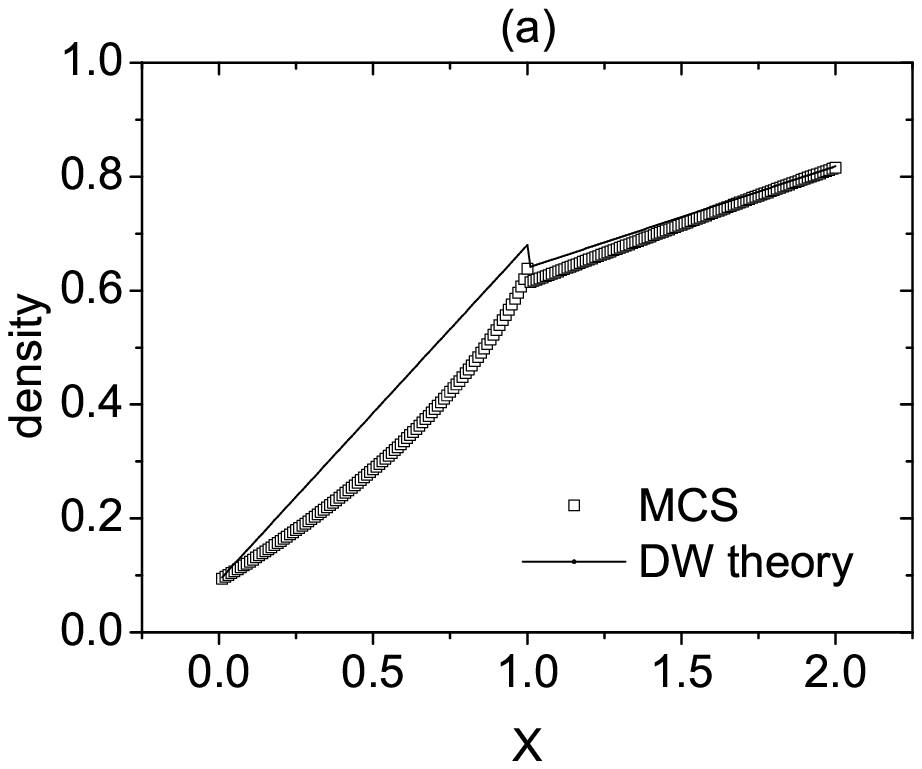} \quad
\includegraphics[width= 2.5 in, height=2 in]{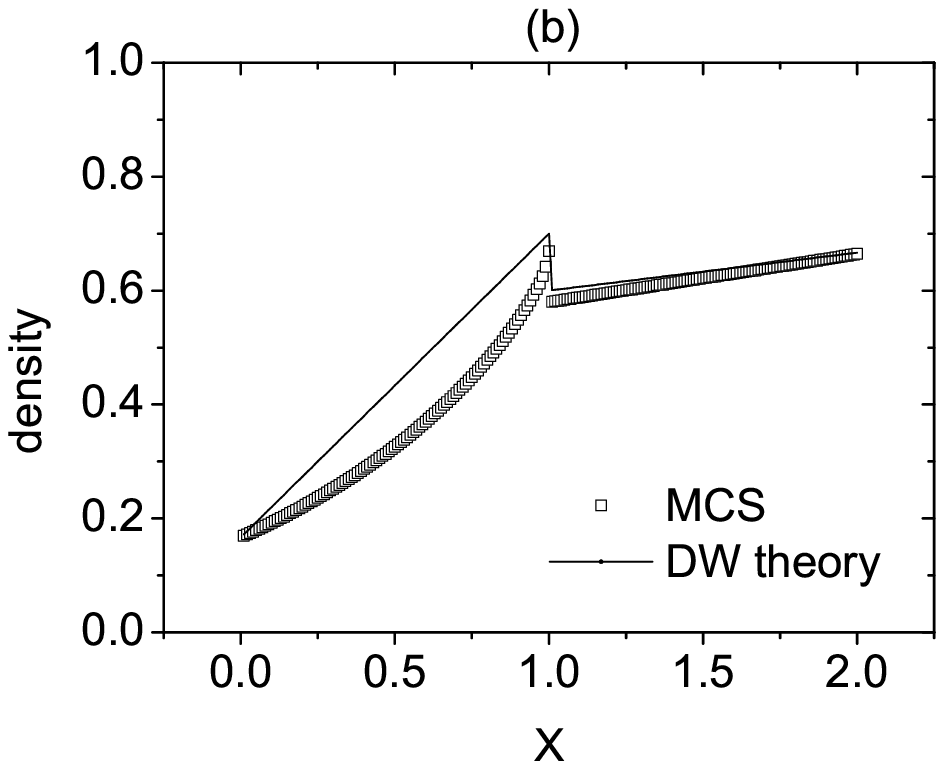} \quad
\includegraphics[width= 2.5 in, height=2 in]{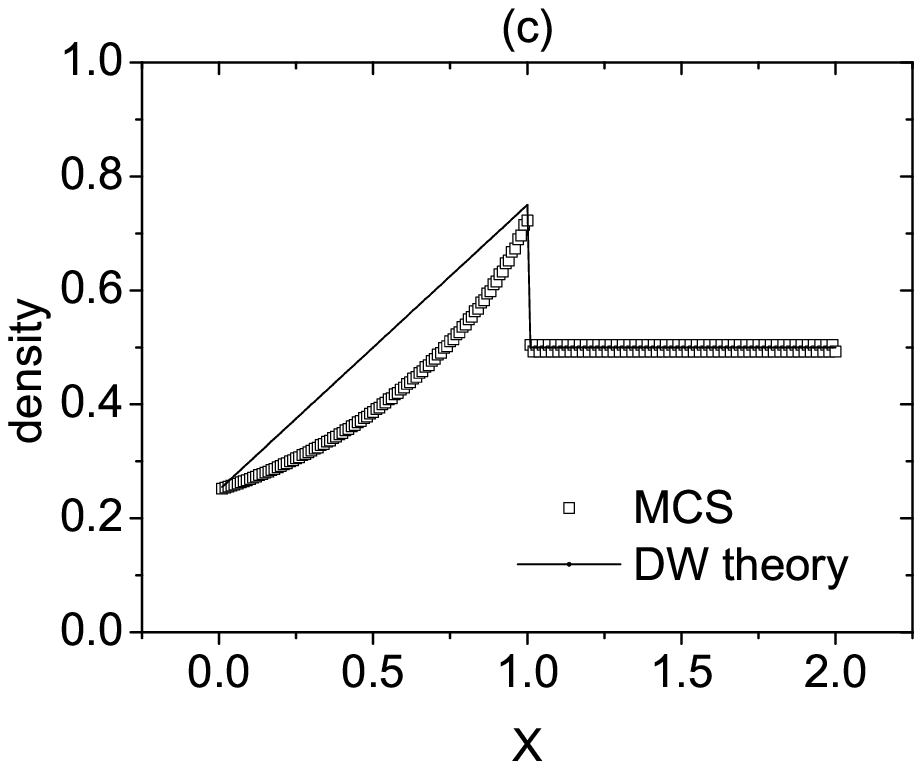} \quad
\caption{Density profiles obtained by the domain wall (DW) theory
and Monte Carlo simulations (MCS) when $m=2$: (a) and (b) for the
phase coexistence line between the (LD, LD) and (HD, HD) phases, (c)
is for the coexistence phases between the (LD, LD), (LD,
HD),(HD,HD)and (HD, MC) phases. The parameters are: (a) $\alpha=0.1
$ and $ \beta=0.222$, (b) $\alpha=0.2$ and $ \beta=0.5$, and (c)
$\alpha=1/3 $ and $ \beta=1.0$.}
\end{center}
\end{figure}

A phenomenological domain wall (DW) theory developed in Section III
is used to calculate the density profiles of phase boundaries such
as the (LD, HD) and (LD, MC) phases (see Fig. 5). The results
obtained from the domain wall theory show an agreement with computer
simulations. When $\alpha$ and $\beta$ both increase and also
maintain $\alpha=\beta/[m+(m-1)\beta]$, the system keeps in the (LD,
HD) phase until $\beta = 1$; the slope of the density profiles for
$x<1$ decreases until the slope reduces to 0.5, while the slope of
the density profiles for $1<x<2$ also decrease until the slope
decreases to 0. For instance, the slope decreases from 0.588 to
0.542 (also see Eq. (32)) when $\alpha$ increases from 0.1 to 0.2
(see Figs. 5(a) and (b)). Finally, the slopes of density profiles of
the subchains become 0.5 and the slope of density profile of the
main chain becomes 0 when $\alpha = 1/3$ and $\beta = 1$ (see Fig.
5(c)). Additionally, Monte Carlo simulations, theoretical
calculations and domain wall theory all show that, when $\alpha =
1/3$ and $\beta = 1$ (i.e., the transition phase between the other
four phases), the main chain is in the maximal current phase.

\begin{figure}[!h]
\begin{center}
\includegraphics[width= 2.5 in, height=2 in]{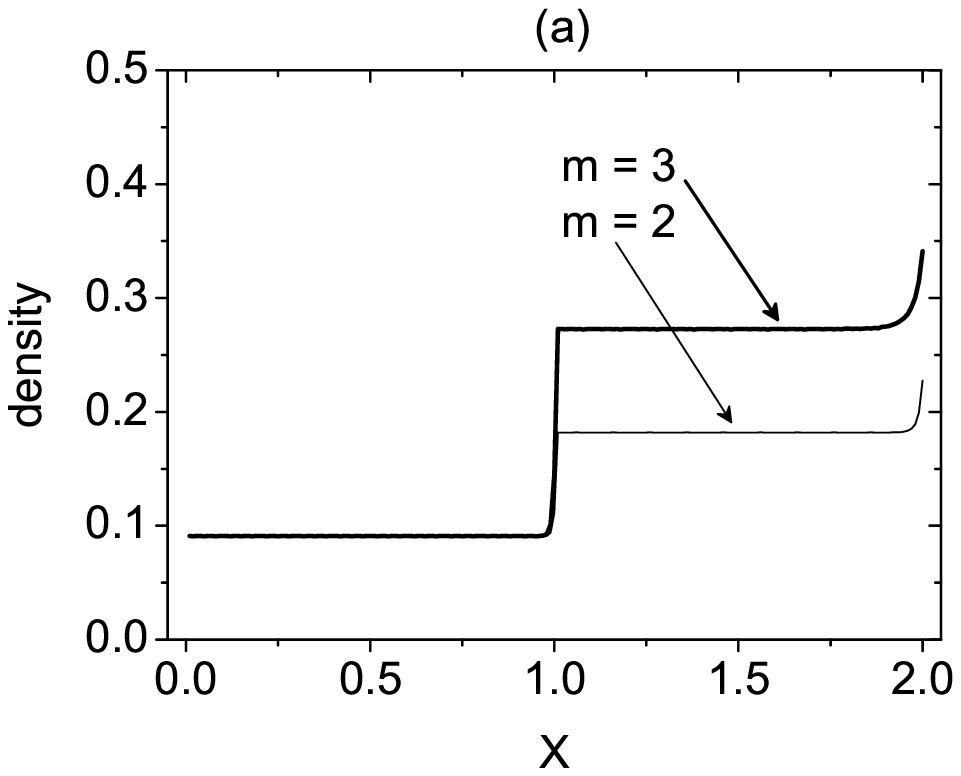} \quad
\includegraphics[width= 2.5 in, height=2 in]{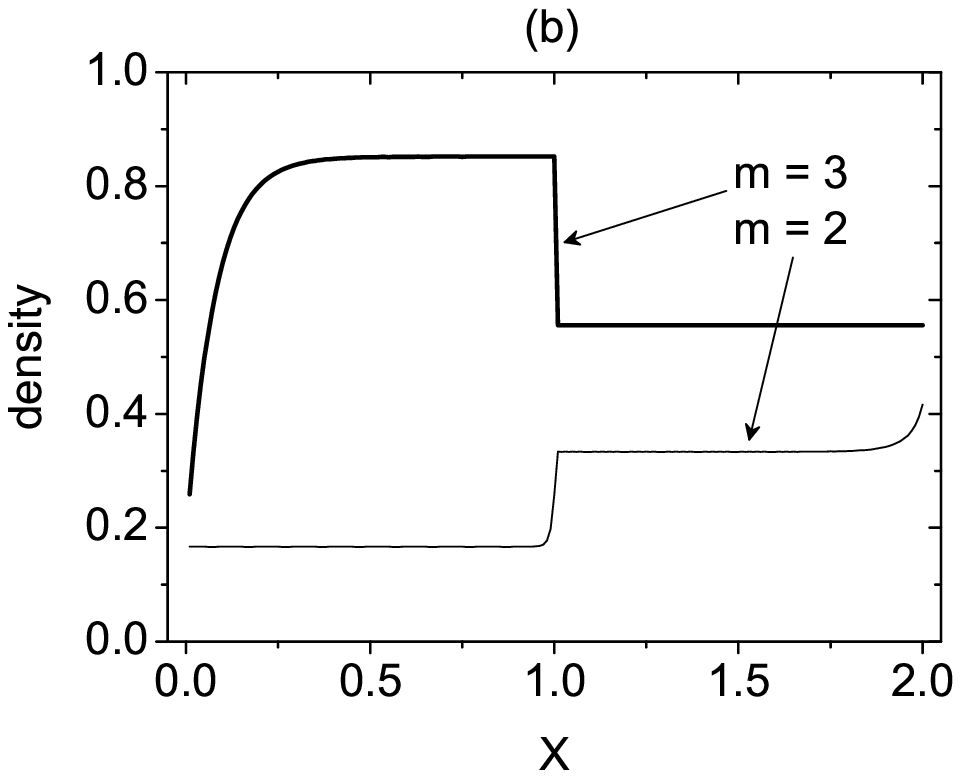} \quad
\includegraphics[width= 2.5 in, height=2 in]{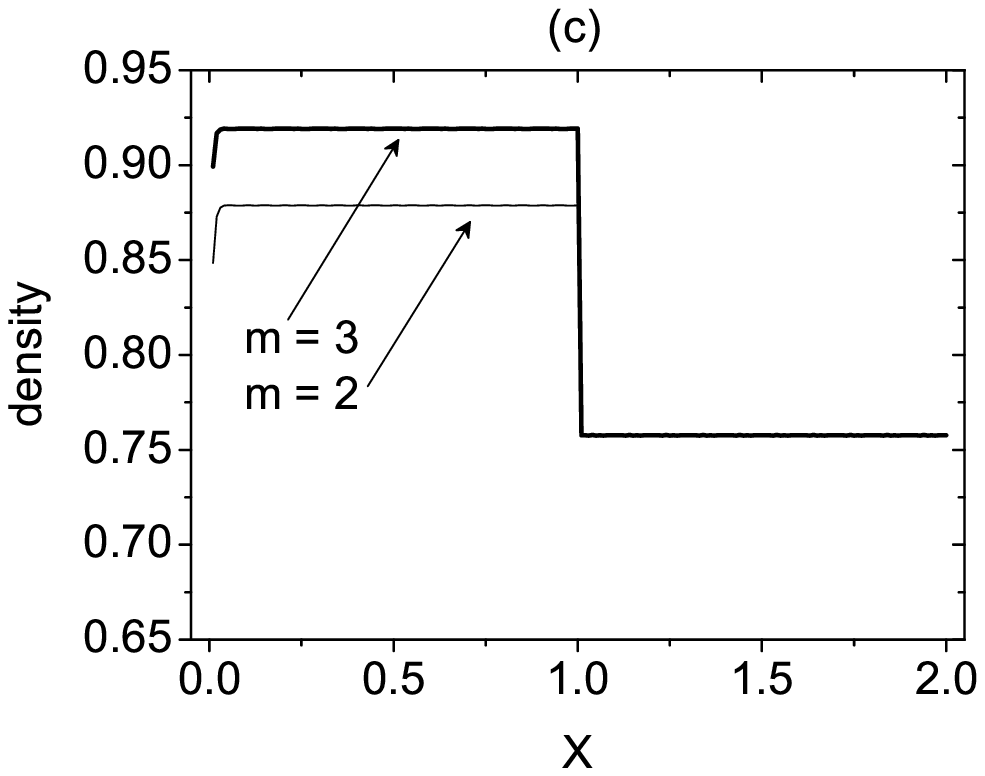} \quad
\includegraphics[width= 2.5 in, height=2 in]{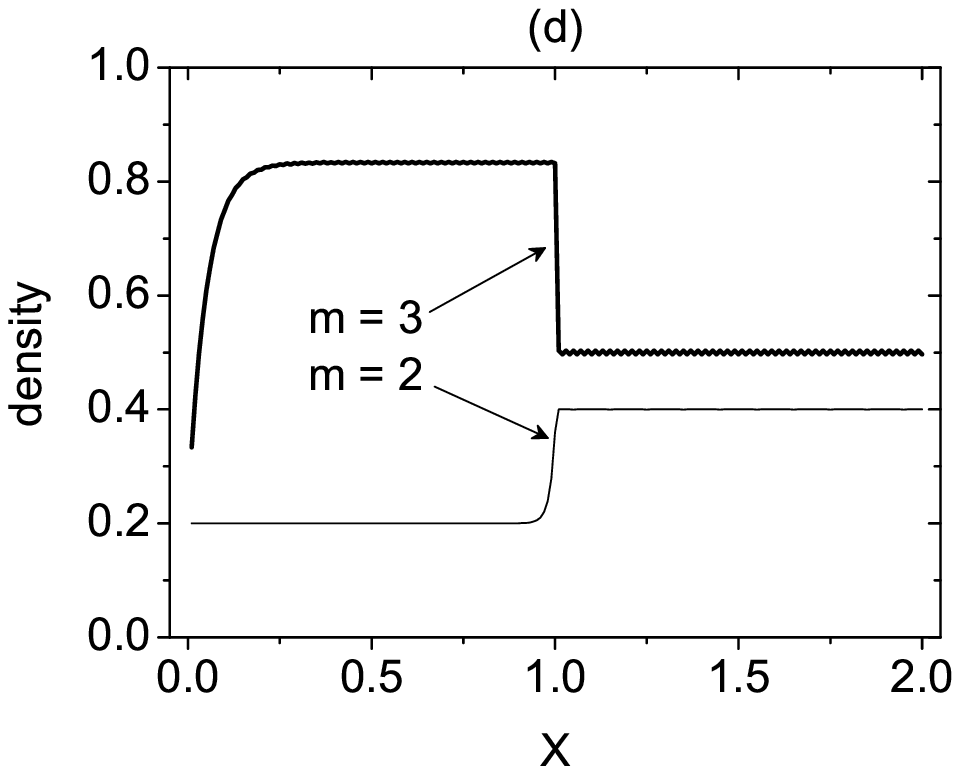} \quad
\caption{\label{fig:m-compare} Density profiles in Monte Carlo
simulations when $m=2$ and $m=3$. The parameters are: (a)
$\alpha=0.1 $ and $ \beta=0.8$, (b) $\alpha=0.2 $ and $ \beta=0.8$,
(c) $\alpha=0.8 $ and $ \beta=0.32$, and (d) $\alpha=0.25 $ and $
\beta=1.0$.}
\end{center}
\end{figure}

Density profiles of the systems for $m=2$ and $m=3$ with the
synchronous update scheme are simulated and compared (see
Fig.~\ref{fig:m-compare}). According to Eq. (7), the phase boundary
between the (LD, LD) and (HD, HD) phases can be described as
$\alpha=\beta/(2+\beta)$ for $m=2$ and $\alpha=\beta/(3+2\beta)$ for
$m=3$. Fig.~\ref{fig:m-compare} shows that both systems are in the
(LD, LD) phase when $\alpha=0.1$ and $\beta=0.8$. However, when
$\alpha$ increases (e.g., $\alpha=0.2$), the system for $m=2$ is
still in the (LD, LD) phase, while the system for $m=3$ is in the
(HD, HD) phase (see Fig.~\ref{fig:m-compare}(b)). This is due to the
phase boundary between the (LD, LD) and (HD, HD) phases moving
towards the left when $m$ increases (see Fig. 2(a)). Density
profiles in the (HD, HD) phase for both $m=2$ and $m=3$ are shown in
Fig.~\ref{fig:m-compare}(c). Compared with
Fig.~\ref{fig:m-compare}(a) and (c), it can be seen that the density
profiles of the subchains of both systems are the same when both
systems in the (LD, LD) phase, while the density profiles of the
main chains of both systems are the same when both systems are in
the (HD, HD) phase. Fig.~\ref{fig:m-compare}(d) illustrates that the
system is in the (LD, LD) phase for $m=2$, while it is in the (HD,
MC) phase for $m=3$.

We can also see the similarities and differences between the phase
diagram of the system with the \emph{synchronous}/\emph{parallel}
update scheme (see Fig. 2(b)) and that of the system with the
\emph{random} update scheme (see Fig. 3 in \cite{PRO05}). One can
see that the structures of the phase diagrams are similar. All have
five phases in their phase diagrams. Also, increasing the number of
subchains (i.e., inputs) only shifts the transition line between the
(LD, LD) phase and the (HD, HD) phase that does not fall on the
boundaries of the phase diagrams. However, the differences in the
phase diagrams include: (i) the (HD, MC) phase region in the phase
diagram of the system with the random update scheme reduces to a
line in that of the system with the synchronous update scheme; and
(ii) the line of the (LD, MC) phase in the phase diagram of the
system with the random update scheme reduces to a point in the phase
diagram of the system with the synchronous update scheme.

\begin{figure}[!h]
\begin{center}
\includegraphics[width= 2.5 in, height=2 in]{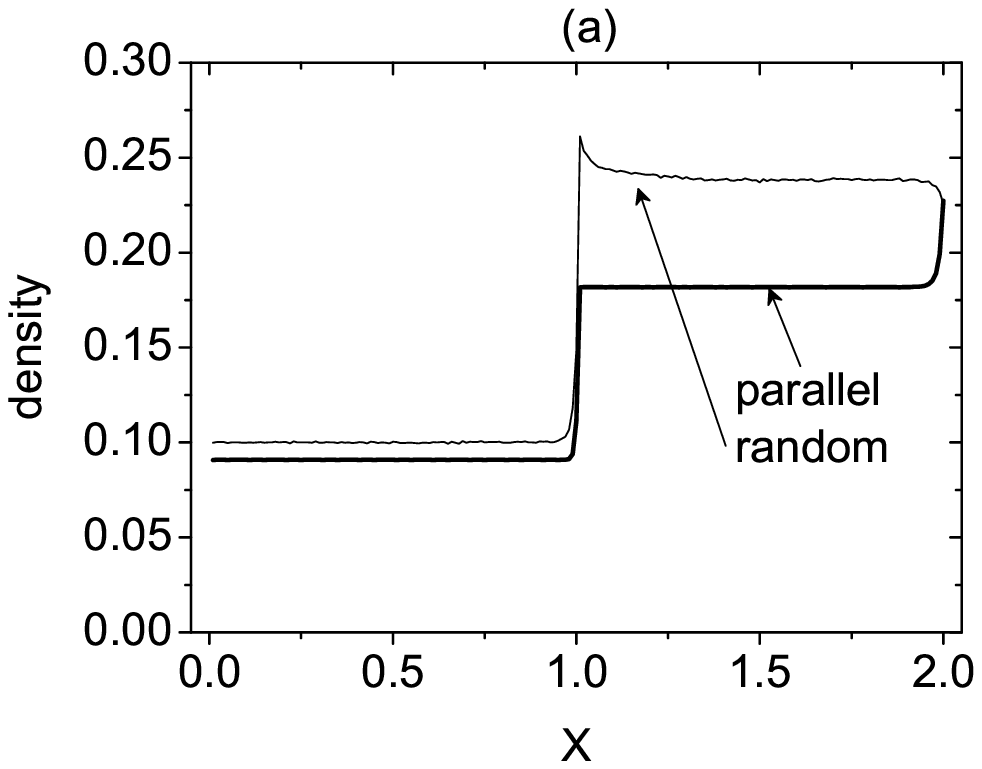} \quad
\includegraphics[width= 2.5 in, height=2 in]{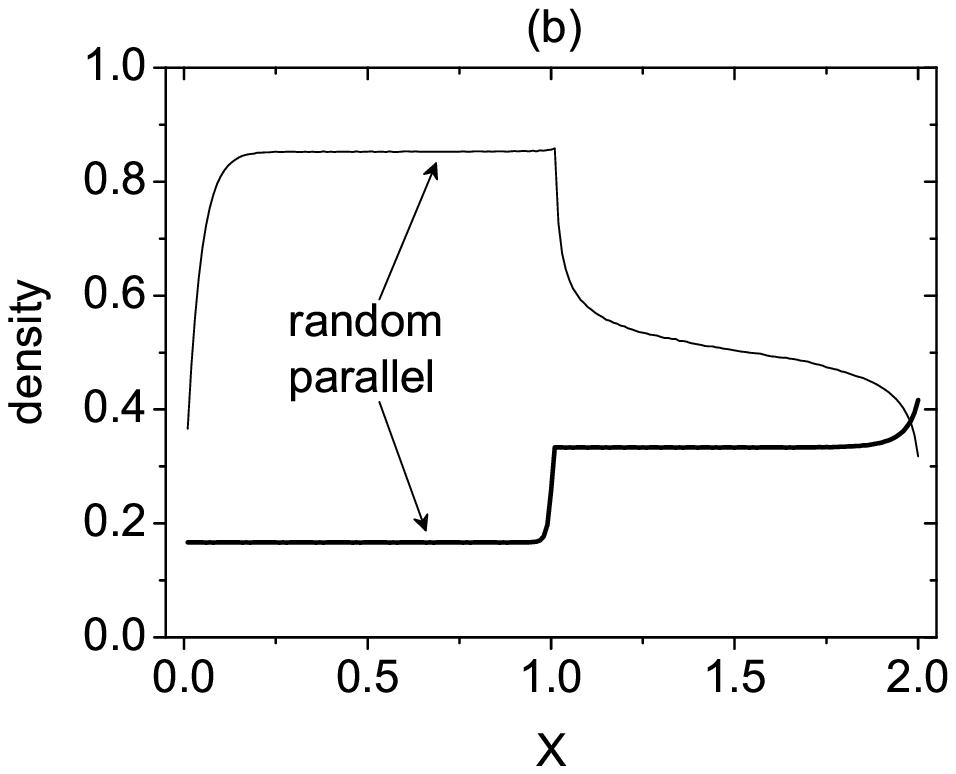} \quad
\includegraphics[width= 2.5 in, height=2 in]{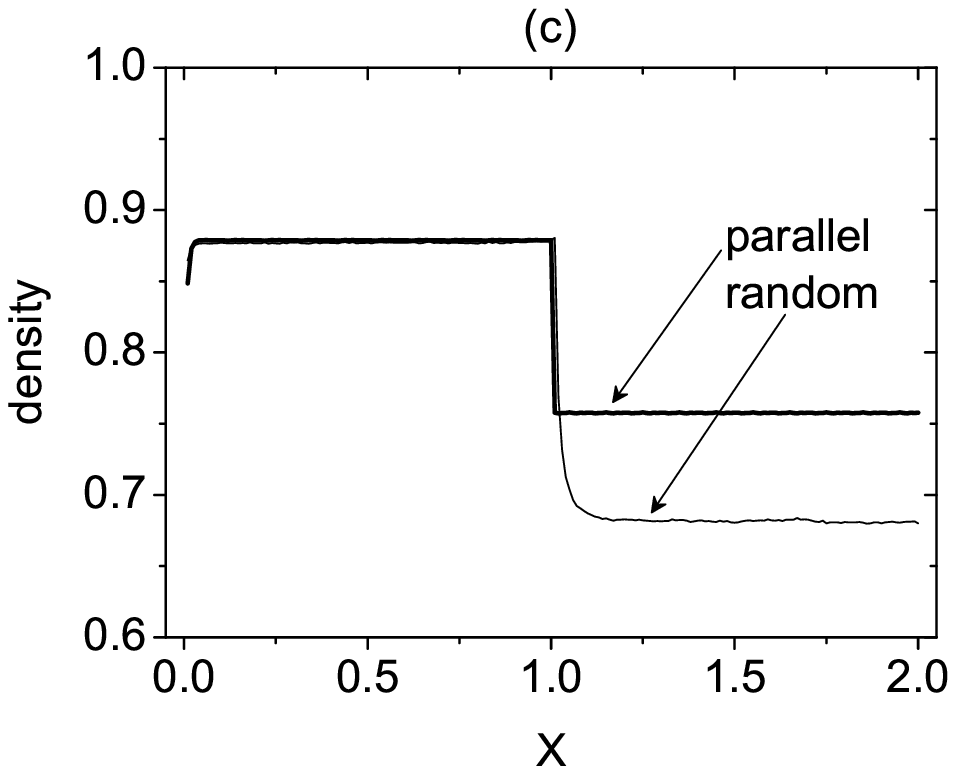} \quad
\includegraphics[width= 2.5 in, height=2 in]{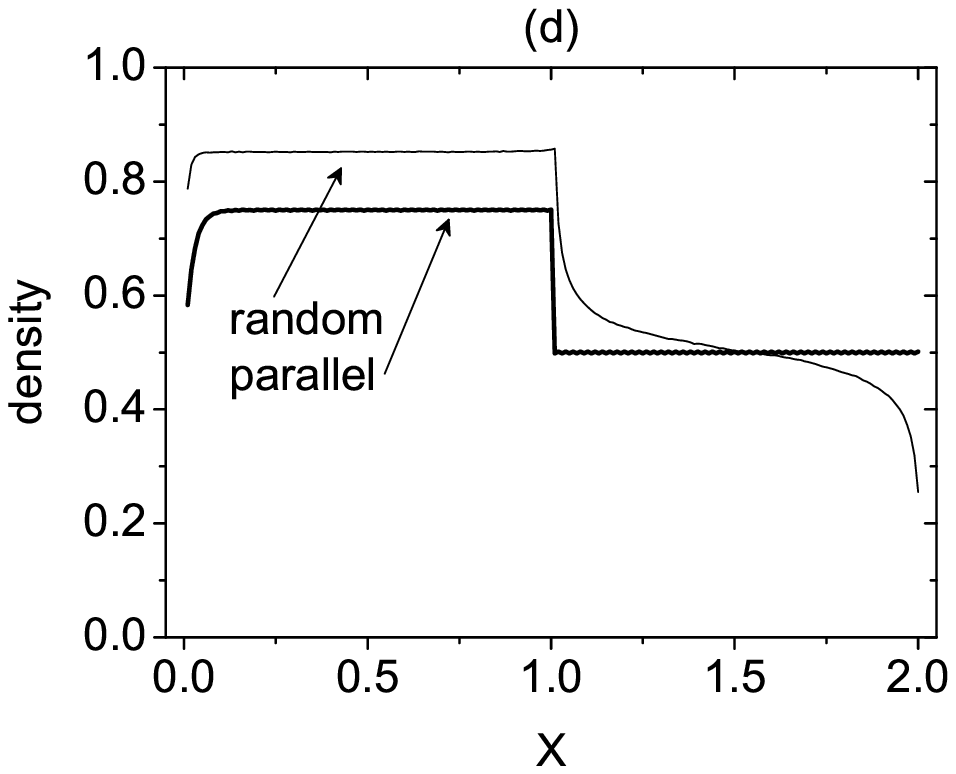} \quad
\caption{\label{fig:r-p-compare} Density profiles in Monte Carlo
simulations when $m=2$ in random and parallel updates. The
parameters are: (a) $\alpha=0.1$ and $\beta=0.8$, (b) $\alpha=0.2$
and $ \beta=0.8$, (c) $\alpha=0.8$ and $ \beta=0.32$, and (d)
$\alpha=0.8$ and $\beta=0.8$.}
\end{center}
\end{figure}

Fig.~\ref{fig:r-p-compare} shows the differences in the density
profiles of the systems with the \emph{synchronous} update scheme
and the system with the \emph{random} update scheme when $m=2$. In
Fig.~\ref{fig:r-p-compare}(a), these two systems are in the (LD, LD)
phase when $\alpha=0.1$ and $\beta=0.8$. When $\alpha$ is increased
to 0.2 and $\beta$ is unchanged, the system with the synchronous
update scheme is still in the (LD, LD) phase, while the phase of the
system with the random update scheme becomes the (HD, MC) phase (see
Fig.~\ref{fig:r-p-compare}(b)). Fig.~\ref{fig:r-p-compare}(c) shows
the system in the (HD, HD) phase in both systems when $\alpha=0.8$
and $\beta=0.32$. With the increase of $\beta$ (e.g.,
$\beta=0.8$),the phase of the system with the random update scheme
changes to the (HD, MC) phase, while it still keeps in the (HD, HD)
phase in the other system.

Note that the system also exhibits a particle-hole symmetry. Since
particles moving forward at junction points with the same priority
is equivalent to holes moving backward at the same priority. Also,
our method can be used to analyze synchronous TASEPs with a
single-input-multi-output (SIMO) junction. Other inhomogeneous
synchronous TASEPs can be investigated in the similar way. For
instance, it would be interesting to study an MISO junction where
these parallel subchains are dynamically different.

\section{\label{sec:Conclusions}Summary and Conclusions}

Multi-input-single-output (MISO) junctions are relevant to many
biological processes as well as vehicular and pedestrian traffic
flow. \emph{Synchronous} totally asymmetric exclusion processes
(TASEPs) with an MISO junction are investigated in this paper. The
theoretical solutions, mean-field approximation, domain wall theory
are developed. Extensive computer simulations are conducted.  Our
theoretical analysis suggests that there are five possible
stationary phases ((LD, LD), (LD, HD), (LD, MC), (HD, HD) and (HD,
MC)). The (LD, HD) phase corresponds to the transition line (when
$\alpha=\beta/[m+(m-1)\beta]$ and $\beta <1$, where $m$ is the
number of subchains) between the (LD, LD) phase and the (HD,HD)
phase. The (LD, MC) phase (when $\alpha= 1/(2m-1)$ and $\beta =1$)
is the transition phase between the (LD, LD), (LD, HD), (HD, HD) and
(HD, MC) phases. Also, the non-equilibrium stationary state,
stationary-state phases and the phase boundary are determined by the
boundary conditions of the system as well as the number of
subchains. The phase boundary moves to the left in the phase diagram
when the number of subchains increases. The density profiles are
simulated, which shows good agreement with theoretical analysis.

We also compare the phase diagrams and density profiles between the
system with the synchronous update scheme and the system with the
random update scheme. The main differences in the phase diagrams
include: (i) the (HD, MC) phase region in the phase diagram of the
system with the random update scheme reduces to a line in that of
the system with the synchronous update scheme; and (ii) the line of
the (LD, MC) phase in the phase diagram of the system with the
random update scheme reduces to a point in the phase diagram of the
system with the synchronous update scheme.

The approach used in this paper can be used directly to analyze
TASEPs with a single-input-multi-output (MIMO) junction in a
parallel updating procedure.

\section*{Acknowledgements}
The authors gratefully acknowledge the comments and suggestions of
the anonymous reviewers, which helped in improving the clarity and
the quality of the paper. R. Wang acknowledges the support of Massey
University Research Fund (2007) and Massey University International
Visitor Research Fund (2007). R. Jiang acknowledges the support of
is supported by National Basic Research Program of China
(No.2006CB705500),the NNSFC under Project No. 10532060, 70601026,
10672160, the CAS President Foundation, the NCET and the FANEDD. We
are grateful to Michele Wagner for proofreading this manuscript.

\end{document}